\documentclass[preprint,12pt,times]{elsarticle}

\usepackage{amssymb,bm}
\usepackage{graphics}
\usepackage{tensor}
\usepackage{amsmath,multicol}
\usepackage{braket}

\journal{}

\usepackage[usenames,dvipsnames]{color}

\begin{document}

\begin{frontmatter}

\title{{\bfseries Peculiarities of quasifission reactions in heavy ion collisions}}

\author[Dubna,Tashkent,KNU]{Avazbek Nasirov}
\ead{nasirov@jinr.ru}

\author[Tashkent,Inha]{Bakhodir Kayumov}

\author[KNU,APCTP,GWU]{Yongseok Oh}
\ead{yohphy@knu.ac.kr}

\address[Dubna]{Bogoliubov Laboratory of Theoretical Physics, JINR, Dubna, Russia}
\address[Tashkent]{Heavy Ion Physics Laboratory, Institute of Nuclear Physics, Tashkent, Uzbekistan}
\address[KNU]{Department of Physics, Kyungpook National University, Daegu 41566, Republic of Korea}
\address[Inha]{Inha University in Tashkent, Uzbekistan}
\address[APCTP]{Asia Pacific Center for Theoretical Physics, Pohang, Gyeongbuk 37673, Republic of Korea}
\address[GWU]{Institute for Nuclear Studies and Department of Physics,
The George Washington University, Washington, DC 20052, USA}

\begin{abstract}
The probability of the formation and decay of a dinuclear system is investigated for a wide range
of relative orbital angular momentum values.
The mass and angular distributions of the quasifission fragments are studied to understand the reaction
mechanisms of the heavy ion collision of $\nuclide[78]{Kr \mbox{ (10 $A$ MeV)}} + \nuclide[40]{Ca}$
within dinuclear system model.
The quasifission products are found to contribute to the mass-symmetric region of the mass distribution
in collisions with a large orbital angular momentum.
The analysis of mass and angular distributions of quasifission fragments shows the possibility of
the $180^\circ$ rotation of the system so that projectile-like products can be observed in the forward
hemisphere with large cross sections, which can explain the phenomenon observed recently in the
ISODEC experiment.
\end{abstract}


\end{frontmatter}


\section{Introduction}

The dynamics of heavy ion collisions can be studied by analysing energy, mass and
angular distributions of the products observed in various reaction channels.
The correlations between these distributions then allow us to extract the information about
the formation and decay of a metastable composite system, if any, such as a dinuclear system (DNS),
during the contact time of interacting nuclei~\cite{NFTA05}.

In heavy ion collisions the capture of the projectile by the target nucleus leads to the formation of
a molecule-like DNS which evolves by changing its charge and mass asymmetries as well as its
shape due to multinucleon transfers.
The excitation energy $E^*_{Z}$ and the angular momentum $L_{Z}$ of the DNS with a
charge asymmetry $Z$ depend on the collision energy $E_{\rm c.m.}$ in the centre-of-mass system
and the initial value of the relative orbital angular momentum determined by the impact parameter $b$ and
the momentum $\bm{P} = \mu \dot{\bm{R}}$ of the collision, i.e.,
$\bm{L} = \bm{\ell} \hbar = \bm{b} \times \bm{P}$, where $\mu$ is the reduced mass of the
initial state.
The angular distribution of the reaction fragments is one of the informative quantities which enable us to
understand the fusion-fission mechanisms of heavy ion collisions.
In addition, studies on the correlations between mass and angular distributions of the fragments in the
full-momentum-transfer reactions are a way to find a method to separate pure fission fragments of the
compound nucleus with a compact shape from quasifission fragments formed by the DNS decay without
the formation of a compound nucleus.
This circumstance shows that the mass and angular momentum distributions of the reaction fragments
are determined by the collision dynamics and behaviour of the DNS formed at the capture stage.

Deep inelastic heavy ion collisions may also induce the formation of a molecule-like DNS but the
full-momentum-transfer reaction does not take place.
In this case the relative motion of the colliding nuclei is not completely damped and the projectile-like
and target-like products go away.
The lifetime of DNS formed in deep inelastic collisions would be shorter than those of capture cases.

Recently, the analysis in Refs.~\cite{ISODEC-14,Schroeder15} on the inverse-kinematics ISODEC
experiment led to a claim of the observation of a new reaction mechanism in the reaction of
$\nuclide[78]{Kr}(\mbox{$E/A =10$~MeV}) + \nuclide[40]{Ca}$.
In this experiment, the energy and angular distributions of the binary reaction products of the collision
have been measured and the velocity and mass distributions were reconstructed.
It is clear that the observed yields of the binary products are related with the deep inelastic collisions,
quasifission and fusion-fission processes.
In these works, the last two processes, where full momentum transfer takes place, were analysed,
and the events with the component of the velocity distribution in the range of
$1.5~\mbox{cm/ns} < {\rm v}_{\rm rel}^{} < 3.5~\mbox{cm/ns}$ peaking at
${\rm v}_{\rm rel}^{} = 2.4$~cm/ns were explored.
The relative velocity is the difference between velocities of the observed fragments,
${\rm v}_{\rm rel}^{} = \left| \overrightarrow{\rm v}_1^{} - \overrightarrow{\rm v}_2^{} \right|$.
The individual fragment velocity vectors ($\overrightarrow{\rm v}_1^{}$, 
$\overrightarrow{\rm v}_2^{}$) and the corresponding
momenta are used to determine the corresponding velocity components, parallel and perpendicular to
the beam, in the rest frame of the emitter, i.e., the fission source.


\begin{figure}[t]
\centering
\includegraphics[width=1.0\columnwidth]{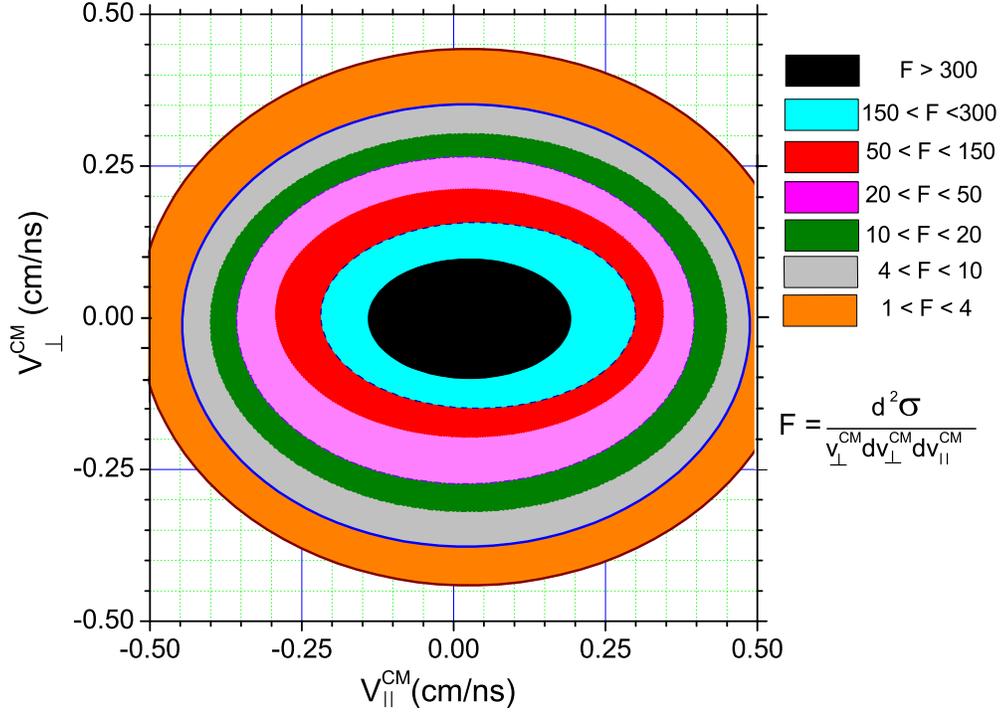}
\caption{(Color online) 
Correlated fragment velocities parallel (${\rm v}_\parallel^{}$) and perpendicular 
(${\rm v}_{\perp}^{}$) to the beam
in the rest frame of the emitter.
The velocity of the centre-of-mass system in the laboratory frame is subtracted and
the bands are built based on Fig.~6 of Ref.~\cite{Schroeder15}.
}
\label{velo12}
\end{figure}

Three phenomena observed in the analysis of the reaction products inspired the authors of
Refs.~\cite{ISODEC-14,Schroeder15} to suggest a new interesting reaction mechanism called
``shock-induced fission following fusion" in central collisions.
These three phenomena are summarised as follows.
\begin{itemize}
\item[i)]
The first observation is that the velocity distribution in the beam direction 
(${\rm v}_{\parallel}$) is slightly deformed whereas the spectrum of transversal relative 
velocity (${\rm v}_{\perp}$) is isotropic.
This can be seen in Fig.~\ref{velo12} where the correlated fragment velocities parallel and
perpendicular to the beam direction in the emitter's rest frame are shown.
The presented results are obtained for the mass-symmetric fission fragments with fission-like
relative velocities and are drawn based on Fig.~6 of Ref.~\cite{Schroeder15} to demonstrate one of
the main arguments of the authors of Refs.~\cite{ISODEC-14,Schroeder15} to state the observation
of the so-called ``shock-induced fission following fusion".
\item[ii)]
The second observation is the unusual properties of the fragment angular distribution
$d\sigma/d\Theta_{\rm HFr}$, which is strongly anisotropic, except for mass-symmetric events, and
not symmetric at $90^{\circ}$ as shown in Fig.~8 of Ref.~\cite{Schroeder15}.
For asymmetric fission events, the heavier fragment is preferentially emitted in the forward direction
in the centre-of-mass system.
For symmetric events, where $|(A_1 - A_2)/(A_1 +A_2)| < 0.1$ with $A_1$ and $A_2$ being
the mass numbers of the fragments,
the distribution is not isotropic and has maxima both at forward and backward angles.
This behaviour indicates a rather strong alignment of the fission axis in the beam direction and
demonstrates the dominant dynamical character of the process.
Clear asymmetric fission events were found to have a tendency that more massive projectile-like
fragments proceed along the beam direction, which seems to be the memory of the initial mass and
velocity distributions.
The observation of the heavy projectile-like fragments beyond the light target-like ones in the reaction of
$\nuclide[78]{Kr}(\mbox{$E/A =10$~MeV}) + \nuclide[40]{Ca}$ seems to be unusual and,
in Refs.~\cite{ISODEC-14,Schroeder15}, the idea of ``shock-induced fission following fusion"
was suggested, which was claimed to occur in central collisions.
\item[iii)]
As shown in Fig.~9 of Ref.~\cite{Schroeder15}, the Galilean-invariant velocity distributions of
$\alpha$-particles emitted from the forward-moving  (mass-symmetric) fission fragments were
observed to be isotropic.
This means that the spin angular momenta of the emitting fragments are negligibly small, which
supports the conclusion that the ``shock-induced fission'' occurs at small initial angular momentum
($\ell\approx 0 \mbox{ -- } 40$).
The possibility of the transparency of the light target \nuclide[40]{Ca} through heavy projectile
\nuclide[78]{Kr} in central fusion-type heavy-ion collisions has been demonstrated by the
presentation of plots of the density contours of projectile- and target-like fragments in central
$\nuclide[78]{Kr\mbox{ (11$A$ MeV)}} + \nuclide[40]{Ca}$ collisions
as a function of time in Fig.~3 in Ref.~\cite{Schroeder15}.
\end{itemize}

In the present work, to understand the data of the ISODEC experiment, we investigate the reaction
of $\nuclide[78]{Kr\mbox{ ($10 A$ MeV)}} + \nuclide[40]{Ca}$ and consider the capture dynamics
of the projectile-target nuclei and the decay process of the DNS formed during the reaction.
This analysis will lead to that the yield of quasifission products can contribute to the heavy products
emitted in the forward hemisphere reported in Ref.~\cite{ISODEC-14} at the relative angular momentum
of $L = \mbox{(60--75)}\hbar$.


\section{Formation of DNS}

The characteristics of the reaction products are related to the properties of the intermediate
DNS formed at the capture stage of the projectile nucleus by the target nucleus.
The DNS lifetime is determined by its excitation energy $E^*_{\rm DNS}$ and the quasifission
(pre-scission) barrier $B_{\rm qf}$ which depend on the orbital angular momentum $L$
at a given value of the centre-of-mass collision energy $E_{\rm c.m.}$~\cite{PRC05}.
The quasifission barrier $B_{\rm qf}$ is found by the minimum and maximum values of the potential
well in the nucleus-nucleus interaction~\cite{NFTA05,PRC05}.
Furthermore, the behaviour of DNS depends on the properties of the interacting nuclei such as shape,
shell effects and orientation angles of the axial symmetry axis relative to the beam direction.
The collision trajectory, rotational angle, angular velocity and moment of inertia for the DNS formed
after capture for a given energy $E_{\rm c.m.}$ and orbital angular momentum $L_0$ are
found by solving the following equations of motion~\cite{NFTA05, AJNM96}:
\begin{eqnarray}
\label{maineq1}
&& \mu(R)\frac{d \dot R}{dt} + \gamma_{R}^{}(R) \dot R(t) = F(R),\\
\label{maineq2}
&& F(R,\alpha_1^{},\alpha_2^{}) = -\frac{\partial V(R,\alpha_1^{},\alpha_2^{})}{\partial R}-
 \dot R^2 \frac{\partial \mu(R)}{\partial R} ,\\
\label{maineq3}
&& \frac{dL}{dt} = \gamma_{\theta}^{}(R) R(t) \left( \dot{\theta} R(t)
- \dot{\theta_1} R_{\rm 1,eff} - \dot{\theta_2} R_{\rm 2,eff}\right), \\
\label{maineq4}
&& \frac{dL_1}{dt} = \gamma_{\theta}^{} (R) \left[ R_{\rm 1,eff} \left( \dot{\theta} R(t) -
\dot{\theta_1} R_{\rm 1,eff} - \dot{\theta_2} R_{\rm 2,eff} \right)
- 2 a \left(R_{\rm 1,eff} \dot{\theta_1^{}} - R_{\rm 2,eff} \dot{\theta_2^{}} \right) \right], \\
\label{maineq5}
&& \frac{dL_2}{dt} = \gamma_{\theta}^{}(R) \left[ R_{\rm 1,eff}  \left( \dot{\theta} R(t) - \dot{\theta_1}
R_{\rm 1,eff} - \dot{\theta_2} R_{\rm 2,eff} \right)
+ 2 a \left(R_{\rm 1,eff} \dot{\theta_1} - R_{\rm 2,eff} \dot{\theta_2} \right) \right],
\\
&& L_0 = L(\dot{\theta}) + L_1(\dot{\theta_1}) + L_2(\dot{\theta_2}) ,\\
&& L(\dot{\theta}) = J_{\rm DNS}(R,\alpha_1^{},\alpha_2^{}) \dot{\theta}, \\
&& L_1(\dot{\theta_1}) = J_1 \dot{\theta_1} , \\
&& L_2(\dot{\theta_2}) = J_2 \dot{\theta_2} , \\
&& E_{\rm rot} = \frac{J_R(R,\alpha_1^{},\alpha_2^{}) \, \dot{\theta}^2}{2}
+ \frac{J_1 \dot{\theta}_1^2}2+\frac{J_2 \dot{\theta}_2^2}2 ,
\end{eqnarray}
where $R(t)$ is the relative distance, $\dot R(t) \equiv dR(r)/dt$ is the corresponding velocity,
$\alpha_1^{}$ and $\alpha_2^{}$ are the orientation angles between the beam direction and
axial symmetry axis of the projectile and the target, respectively, $J_R$ and $\dot\theta$, $J_1$ and
$\dot\theta_1$, $J_2$ and $\dot\theta_2$ are the moments of inertia and angular velocities of the DNS
and its fragments, respectively.
We also defined $R_{\rm 1,eff} = R_1 + a$ and $R_{\rm 2,eff} = R_2 + a$, where
$R_1$ and $R_2$ are the radius of interacting nuclei with $a = 0.54$~fm~\cite{NFTA05}.
Here, $L_0$ and $E_{\rm rot}$ are determined by the initial condition.
The  moment of inertia of DNS is then calculated by the rigid-body approximation as
\begin{eqnarray}
\label{DNSJ}
J_{\rm DNS}(\alpha_1^{},\alpha_2^{};R) = \mu(R) \, R^2(\alpha_1^{},\alpha_2^{})
+ J_1 + J_2,
\end{eqnarray}
where $R(\alpha_1^{},\alpha_2^{})$ is the distance between the centres of nuclei at their given
mutual orientations.

The moment of inertia of the axially deformed nucleus for the rotation around the axis perpendicular
to its axial symmetry is calculated as
\begin{eqnarray}
\label{fragJ}
J_i=\frac{M_i}{5}
\left(R_{i, \perp}^2 + R_{i, \parallel}^2 \right),
\end{eqnarray}
for $i = 1, 2$, where $M_i$ is the mass of the nucleus.
Here, $R_{\perp}(\beta_2^{})$ and $R_{\parallel}(\beta_2^{})$ are the nucleus axes which are
perpendicular and parallel to the symmetry axis, respectively, and explicitly they are written as
\begin{eqnarray}
R_{\perp}(\beta_2^{}) &=& R_0 \left[ 1 + \beta_2^{} Y_{20}
\left( \frac{\pi}{2} \right) \right],
\nonumber\\
R_{\parallel}(\beta_2^{}) &=& R_0 \left[ 1 + \beta_2^{} Y_{20}(0) \right],
\end{eqnarray}
where $R_0$ is the spherical equivalent radius and $\beta_2^{}$ is the parameter of the
quadrupole deformation.

The nucleus-nucleus potential $V(\ell,\{\alpha_i\};R)$ consists of three parts as
\begin{eqnarray}
V(\ell,\{\alpha_i\};R) = V_{\rm Coul}(\{\alpha_i\};R) + V_{\rm nucl}(\{\alpha_i\};R)
+ V_{\rm rot}(\ell,\{\alpha_i\};R),
\end{eqnarray}
where $V_{\rm Coul}$, $V_{\rm nucl}$ and $V_{\rm rot}$ are the Coulomb, nuclear and rotational
potentials, respectively.
We refer to Ref.~\cite{FGLM04} and Appendix~A of Ref.~\cite{NFTA05} for the detailed expressions
of these potentials in terms of the orientation angles of the symmetry axis of the colliding nuclei.

In Eq.~(\ref{maineq1}), $\mu(R)$ is the inertial mass of the radial motion, $\gamma_{R}^{}$ and
$\gamma_{\theta}^{}$ are, respectively, the friction coefficients for the relative motion along $R$
and the tangential direction when two nuclei roll on each other's surfaces.
These kinetic coefficients are calculated microscopically from the coupling term between
the collective relative motion and single-particle excitations of nucleons in the interacting nuclei
by estimating the evolution of the coupling term between the relative motion of nuclei and the nuclear
motion inside nuclei.
More details can be found, e.g., in Refs.~\cite{NFTA05,AJNM96}, where the friction coefficients
are calculated as
\begin{eqnarray}
\label{RadFric}
\gamma_{R}^{} (R(t)) &=& \sum_{i,i'}\left| \frac{\partial V_{i i'}(R(t))}{\partial R}\right|^{\,2}
B^{\,(1)}_{ii'}(t) \, , \\
\label{TanFric}
\gamma_{\theta}^{} (R(t)) &=& \frac {1}{R^2} \sum_{i,i'}\left|
\frac{\partial V_{i i'}(R(t))}{\partial \theta}\right|^{\,2} B^{\,(1)}_{i i'}(t) \, .
\end{eqnarray}
The  dynamic contribution $\delta V (R(t))$ to the nucleus-nucleus potential
$V (R(t)) = V_0(R(t)) + \delta V (R(t))$ is found through the collision trajectory as
\begin{equation}
\delta V (R(t))    = \sum_{i,i'} \left|\frac{\partial V_{i i'}(R(t))}{\partial R} \right|^{\,2}
B^{(0)}_{ii'}(t) \, .
\end{equation}

\begin{figure*}[t]
\centering
\includegraphics[width=\textwidth]{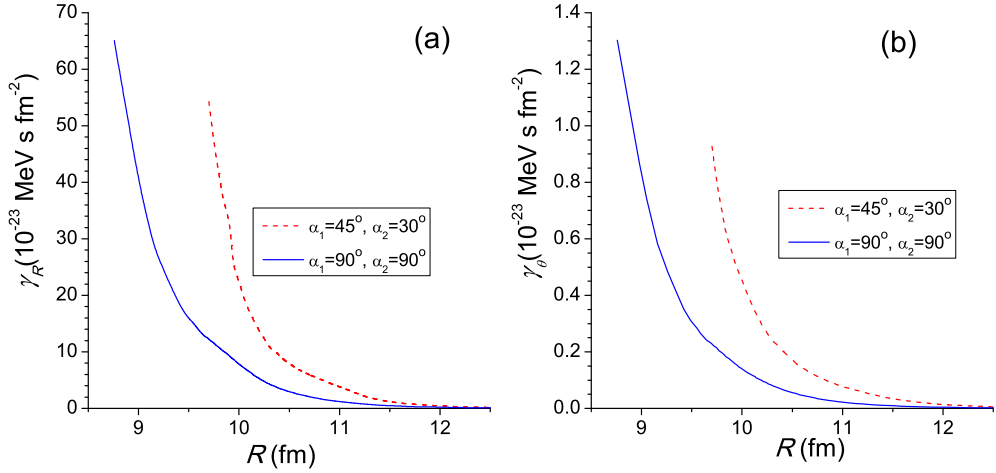}
\vspace*{-3.5 cm}
\caption{
(a) The radial friction coefficient and (b) tangential friction coefficient calculated by
Eqs.~(\ref{RadFric}) and (\ref{TanFric}), respectively, for the $\nuclide[78]{Kr} + \nuclide[40]{Ca}$
reaction at different orientation angles of the axial symmetry axis of the projectile and target nucleus.
The initial value of the orbital angular momentum is taken to be $L=70 \hbar$.
}
\label{FricCoef}
\end{figure*}

The dependence of the radial and tangential friction coefficients on the  orientation angle of the axial
symmetry axis of the projectile and target nucleus is demonstrated in Figs.~\ref{FricCoef}(a) and
\ref{FricCoef}(b), respectively, for the $\nuclide[78]{Kr} + \nuclide[40]{Ca}$ reaction.
These results are obtained for the initial value of the orbital angular momentum $L=70 \hbar$.

The dynamic correction $\delta \mu(R(t))$ of the reduced mass is defined by
$\mu(R(t))=\mu_0+\delta \mu(R(t))$ and is calculated as
\begin{eqnarray}
\delta \mu (R(t)) = \sum_{i,i'}\left|\frac{\partial V_{i i'}(R(t))}{\partial R}\right|^{\,2}
B^{(2)}_{ii'}(t)
- \mu_0^{} \frac{2}{A_{\rm CN}} \int\frac{\rho^{(0)}_1(r - r_1^{})
\rho^{(0)}_2(r - r_2^{})}{\rho^{(0)}_1 (r - r_1^{}) + \rho^{(0)}_2 (r-r_2^{})} d^3r,
\end{eqnarray}
where $\mu_0^{} = m A_1 A_2/A_{\rm CN}$, $m$ is the nucleon mass, $A_1$ ($A_2$) and
$\rho^{(0)}_1$ ($\rho^{(0)}_2$) are the mass number and nucleon density function of the projectile-like
(target-like) fragment of DNS,  respectively, and $A_{\rm CN}=A_1+A_2$ is the mass number of the
compound nucleus.
The time-dependent function $B^{\,(n)}_{i i'}(t)$ is given by
\begin{eqnarray}
\label{Bik}
B_{ik}^{(n)}(t) = \frac{2}{\hbar}\int_{0}^{t}dt' (t-t')^n
\exp\left(\frac{t'-t}{\tau_{ik}^{}}\right) [n_k^{}(t') - n_i^{}(t')]
\sin \left[\omega_{ik}^{} \left(R(t')\right)(t-t')\right],
\end{eqnarray}
with $\hbar\omega_{ik}^{} = \varepsilon_i^{} - \varepsilon_k^{}$ and
$\tau_{i k}^{} = \tau_i^{} \tau_k^{} / (\tau_i^{} + \tau_k^{})$.
Here $n_i^{}$ and $\varepsilon_i^{}$ are the occupation number and energy of a single-particle
state of the DNS fragments and $V_{ii'}$ are the matrix elements of the nucleon exchange between
fragments and particle-hole excitations in the fragments. 
$\tilde\varepsilon_{P_Z}^{}$ and $\tilde\varepsilon_{T_Z}^{}$ are the perturbed energies of
single-particle states: $\tilde\varepsilon_i^{}=\varepsilon_i^{}+V_{ii}$, where $V_{ii}$ is the diagonal
elements of the matrix $V_{ii'}$~\cite{ANAJ94,AJN94}.
The details can be found in Refs.~\cite{AJNM96,ANAJ94}.

The lifetime of the quasiparticle excitations in the single-particle state $i$ of the nucleus is represented by $\tau_i^{}$.
It determines the damping of a single-particle motion and is calculated from the quantum liquid
theory~\cite{PN} and the effective nucleon-nucleon forces~\cite{Migdal} as
\begin{eqnarray}
\label{deftau}
\frac {1}{\tau^{(\alpha)}_i} &=&
\frac{\sqrt{2} \pi}{32 \hbar \varepsilon^{(\alpha)}_{F_K}}
\, \biggl[ \left(f_K - g \right)^2 + \frac{1}{2} \left( f_K + g \right)^2 \biggr]
\nonumber\\ && \mbox{} \times
\biggl[ \left( \pi \, T_K \right)^2 + \left( \tilde{\varepsilon}_i^{} - \lambda^{(\alpha)}_K \right)^2 \biggr]
\left[ 1 + \exp \left( \frac{\lambda^{(\alpha)}_K - \tilde{\varepsilon}_i^{}}{T_K} \right) \right]^{-1},
\end{eqnarray}
where
\begin{equation}
T_K = 3.46 \sqrt{\frac{E_K^*}{\braket{A_K}}}
\end{equation}
is the effective temperature determined by the amount of intrinsic excitation energy
$E_K^{*} = E_K^{*(Z)}+E_K^{*(N)}$ and the mass number
$\braket{A_K(t)}$ with $\braket{A_K(t)} = \braket{Z_K(t)} + \braket{N_K(t)}$.
In addition, $\lambda _K^{(\alpha)}$ and $E_K^{*(\alpha )}$ are the chemical potential and
intrinsic excitation energy for the proton (when $\alpha =Z$) and neutron (when $\alpha =N$)
subsystem of the nucleus $K$, where $K=1$ for the projectile nucleus and $K = 2$ for the
target nucleus, respectively.
Furthermore, by considering the finite size of the nuclei and the difference between the numbers
of neutrons and protons, the Fermi energies are written as~\cite{Migdal}
\begin{eqnarray}
\varepsilon^{(Z)}_{F_K} &=& \varepsilon_F^{}\, \left[ 1 - \frac{2}{3} \left( 1 + 2 f'_K \right)
\frac{\braket{N_K} - \braket{Z_K}}{\braket{A_K}} \right],\\
\varepsilon^{(N)}_{F_K} &=& \varepsilon_F^{}\,
\left[ 1 + \frac{2}{3} \left( 1 + 2 f'_K \right) \frac{\braket{N_K} - \braket{Z_K}}{\braket{A_K}} \right],
\end{eqnarray}
where $\varepsilon_F^{} = 37$~MeV and
\begin{eqnarray}
f_K &=& f_{\rm in} - \frac{2}{\braket{A_K}^{1/3}}(f_{\rm in}-f_{\rm ex}),\\
f'_K &=& f'_{\rm in} - \frac{2}{ \braket{A_K}^{1/3}}(f'_{\rm in}-f'_{\rm ex})
\end{eqnarray}
with $f_{\rm in} = 0.09$, $f_{\rm in}' = 0.42$, $f_{\rm ex} = -2.59$ and $f_{\rm ex}'  = 0.54$.
The coupling constant of the effective nucleon-nucleon interaction is given by $g = 0.7$.


\section{Decay of DNS and angular distribution}

Once the angular momentum $L_{\rm DNS}$ and moment of inertia $J_{\rm DNS}$ of
the dinuclear system are known, its angular velocity is obtained as
$\Omega_{\rm DNS} = L_{\rm DNS} / J_{\rm DNS}$.
To find the angular distribution of the quasifission fragments, we estimate the rotational angle
$\theta_{\rm DNS}$ at the break-up of the system as
\begin{equation}
\theta_{\rm DNS} = \theta_{\rm in} + \Omega_{\rm DNS} \cdot \tau_{\rm DNS}^{},
\label{tetaDNS}
\end{equation}
where $\theta_{\rm in}$ is determined by the dynamical calculations of Eqs.~(\ref{maineq3})
and (\ref{maineq4}) for the entrance channel of the reaction, i.e., at the capture stage.
The value of $\theta_{\rm in}$ depends on the angular momentum and orientation angles $\alpha_1^{}$
and $\alpha_2^{}$ of the symmetric axis of the colliding nuclei at a given $E_{\rm c.m.}$.
The lifetime of the DNS configuration $\tau_{\rm DNS}^{}$ with $Z = Z_1$ and
$Z_2 = Z_{\rm CN} - Z$, where $Z_{\rm CN}$ is the charge number of the compound nucleus,
is determined by the quasifission barrier $B_{\rm qf}$ and the excitation energy $E^*_Z$ for
given values of beam energy and angular momentum $\ell$ through
\begin{equation}
\tau_{\rm DNS}^{}  =
\frac{\hbar}{\Lambda^{\rm qf}_{Z}},
\end{equation}
where the decay width of the DNS is given by~\cite{NMUF07,SJ87}
\begin{eqnarray}
\Lambda^{\rm qf}_{Z} & =&
\frac{\sqrt{\gamma^2_R(R_m)/[2\mu(R_m)]^2 + \omega^2_{\rm qf}(\ell)}
- \gamma_R(R_m)}{2\mu(R_m)}
\nonumber\\
&& \mbox{}\times
\left[\frac{K_{\rm rot} \,\omega_m^{} \exp\left(-B_{\rm qf}(\ell)/T_Z(\ell)\right)}
{2\pi\omega_{\rm qf}^{}(\ell)} \right]
\label{DNS-LF}
\end{eqnarray}
with $R_m$ being the distance between the centres of mass of the DNS fragments
corresponding to the minimum value of the potential well of the nucleus-nucleus interaction.

Equation~(\ref{DNS-LF}) shows that the DNS decay width $\Lambda^{\rm qf}_{Z}$ is proportional to
$K_{\rm rot}$, the collective enhancement factor of the rotational motion to the level density.
Assuming that the DNS is a good rotator, it is estimated as~\cite{JDCI98}
\begin{equation}
K_{\rm rot}(E^*_{Z}) =
\left\{ \begin{array}{ll}
(\sigma_{\bot}^2-1)f(E^*_{Z}) + 1  \qquad & \mbox{if } \sigma_{\bot}^2 > 1\, , \\[5pt]
1 & \mbox{if } \sigma_{\bot}^2 \le 1 \,,  \end{array}
\right.
\end{equation}
where  $\sigma_{\bot}^2 = J_{\rm DNS} T/\hbar^2$,
$f(E) = \left(1+\exp[(E-E_{\rm cr})/d_{\rm cr}] \right)^{-1}$,
$E_{\rm cr}=120\, \tilde{\beta}_2^2 \, A^{1/3}$~MeV and
$d_{\rm cr}=1400\, \tilde{\beta}_2^2 \, A^{-2/3}$.
The effective quadrupole deformation for the dinuclear system is represented by $\tilde{\beta}_2$,
which can be obtained from the value of $J_{\rm DNS}$.
We refer the details to Ref.~\cite{JDCI98}.

\begin{figure}[t]
\centering
\includegraphics[width=1.0\columnwidth]{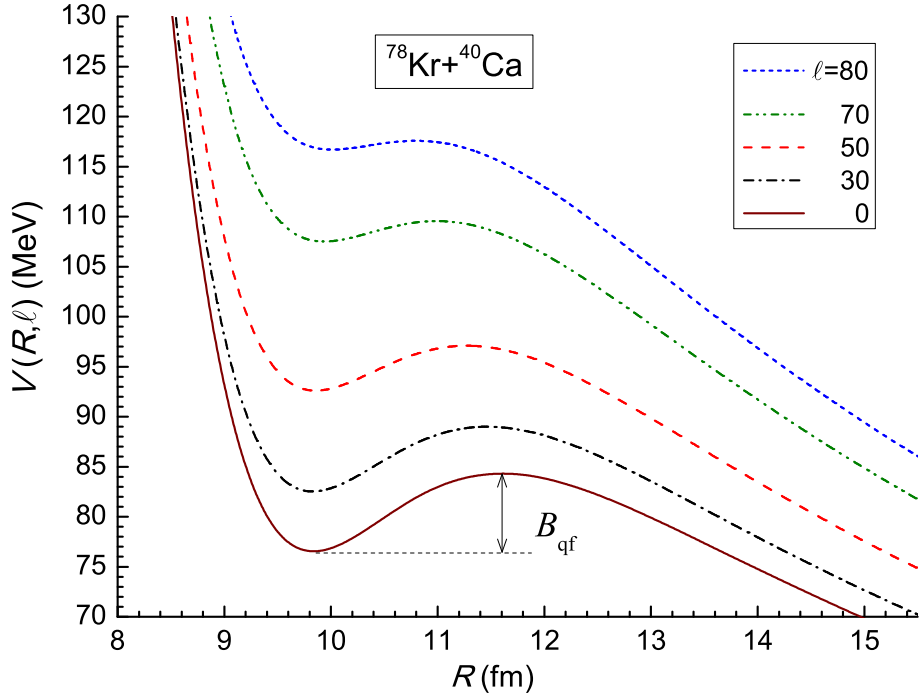}
\caption{(Color online) 
Nucleus-nucleus interaction potential $V(Z,\ell)$ as a function of the distance between the
centres-of-mass of the colliding nuclei calculated for the $\nuclide[78]{Kr} + \nuclide[40]{Ca}$ reaction
at different values of the orbital angular momentum $L = \ell \hbar$.
The quasifission barrier $B_{\rm qf}$ is shown for the potential well
calculated at $L=0$.
The values of $V(R)$ are obtained for $\alpha_1^{} =45^\circ$ and $\alpha_2^{}=15^\circ$ of
the orientation angles of the axial symmetry of the nuclei relative to the beam direction.  }
\label{PotKrCa}
\end{figure}

The frequencies $\omega_m^{}$ and $\omega_{\rm qf}^{}$ are found, respectively, by the harmonic
oscillator approximation to the nucleus-nucleus potential $V(R)$ on the bottom of its pocket placed at
$R_m$ and  on the top of the pre-scission barrier placed at $R_{\rm qf}$, which leads to
\begin{eqnarray}
\omega_m^2&=&\mu_{\rm qf}^{-1}
\left|\frac{\partial^2 V(R)}{\partial R^2}\right|_{R=R_m} ,\\
\omega_{\rm qf}^2&=&\mu_{\rm qf}^{-1}
\left|\frac{\partial^2 V(R)}{\partial R^2}\right|_{R=R_{\rm qf}}.
\end{eqnarray}
The temperature of the DNS consisting of fragments with charge numbers $Z$ and
$Z_{\rm CN}-Z$ is given by
\begin{equation}
\label{Tdns} T_Z = \sqrt{8 E^*_{Z}(\ell)/(A_P+A_T)},
\end{equation}
where $A_P$ and $A_T$ are the mass numbers of the projectile and target nuclei, respectively.
The excitation energy $E^*_Z(\ell)$ of the DNS is determined by the initial beam
energy and the minimum of the potential energy as
\begin{equation}
\label{Edns} E^*_{Z}(\ell) = E_{\rm c.m.}-V(Z,\ell,R_m)+\Delta Q_{\rm gg}(Z),
\end{equation}
where  $V(Z,\ell,R_m)$ is the minimum value of the potential well $V(Z,\ell,R)$
for a given value of $Z$
and $\Delta Q_{\rm gg}(Z)$ is included to take into account the change of the intrinsic energy of the
DNS due to the nucleon transitions during its evolution along the mass and charge asymmetry axes.
The quasifission barrier $B_{\rm qf}$ is determined by the depth of the potential well of the
nucleus-nucleus interaction $V(Z, \ell, R)$ as illustrated in Fig.~\ref{PotKrCa}.
The dependence of $V(Z, \ell,R)$ on the angular momentum of collision is given in Fig.~\ref{PotKrCa}
for the $\nuclide[78]{Kr} + \nuclide[40]{Ca}$ reaction.

The probability of the yield of decay fragment with the charge number $Z$ at time $t$ is then
estimated by
\begin{equation}\label{yield}
Y_{Z}^{}(E^*_Z,\ell,t) = P_{Z}^{} (E^*_Z,\ell,t) \Lambda^{\rm qf}_{Z},
\end{equation}
where $P_Z^{}(E^*_{Z},\ell,t)$ is the probability of population of the
configuration $(Z, Z_{\rm CN}-Z)$ for a given set of $E^*_{Z}$ and $\ell$.
The evolution of the DNS charge asymmetry $P_Z$ is calculated by the
transport master equation:
\begin{eqnarray}
\label{massdec}
\frac{\partial}{\partial t} P_{Z}^{}(E^*_Z,\ell,t) &=&
\Delta^{(-)}_{Z+1} P_{Z+1}^{} (E^*_Z,\ell,t)+
\Delta^{(+)}_{Z-1}  P_{Z-1}^{} (E^*_Z,\ell,t)
\nonumber\\ && \mbox{}
- \left(\Delta^{(-)}_{Z}+\Delta^{(+)}_{Z}+\Lambda^{\rm qf}_{Z} \right)
P_{Z}^{}(E^*_Z,\ell,t)
\end{eqnarray}
for $Z=2,3, \dots , Z_{\rm CN}-2$.
Here, the transition coefficients of multinucleon transfer  are calculated as~\cite{JMN86}
\begin{eqnarray}
\label{delt} \Delta^{(\pm)}_{Z} =
\frac{4}{\Delta t}
\sum\limits_{P,T}|g^{(Z)}_{PT}|^2 \ n^{(Z)}_{T,P}(t) \
\left( 1 - n^{(Z)}_{P,T}(t) \right)
 \frac{\sin^2 [ \Delta t(\tilde\varepsilon_{P_Z}^{}-
 \tilde\varepsilon_{T_Z}^{})/2\hbar ]}
 {(\tilde\varepsilon_{P_Z}^{} - \tilde\varepsilon_{T_Z}^{})^2},
\end{eqnarray}
where the matrix elements $\{ g_{PT}^{} \}$ describe one-nucleon exchange between the DNS nuclei
and their values can be calculated microscopically.
In the present work, we follow Ref.~\cite{AJN92} and estimate these values
with $\Delta t=10^{-22} \mbox{ s} \ll \,t_{\rm int}$.
A non-equilibrium distribution of the excitation energy between the fragments was used in the
calculation of the single-particle occupation numbers $n_{P}^{(Z)}$ and $n_{T}^{(Z)}$ following
Ref.~\cite{AJN94}.

In Eq.~(\ref{massdec}), $\Lambda^{\rm qf}_{Z}$ is the Kramer's rate for the decay probability of
DNS into two fragments with charge numbers $Z$ and $Z_{\rm CN}-Z$~\cite{AAS03},
which is proportional to $\exp [-B_{\rm qf}(Z)/(kT) ]$.
Equation~(\ref{massdec}) with the coefficients (\ref{delt}) and the initial condition
$P_{Z}(E^*,0)=\delta_{Z,Z_P}^{}$ is solved numerically and the primary mass and charge
distributions are found for a given interaction time $t_{\rm int} = 5 \times 10^{-21}$~s~\cite{TBDG85}.


\section{Results and discussion}

\begin{figure}[t]
\centering
\includegraphics[width=1.25\columnwidth]{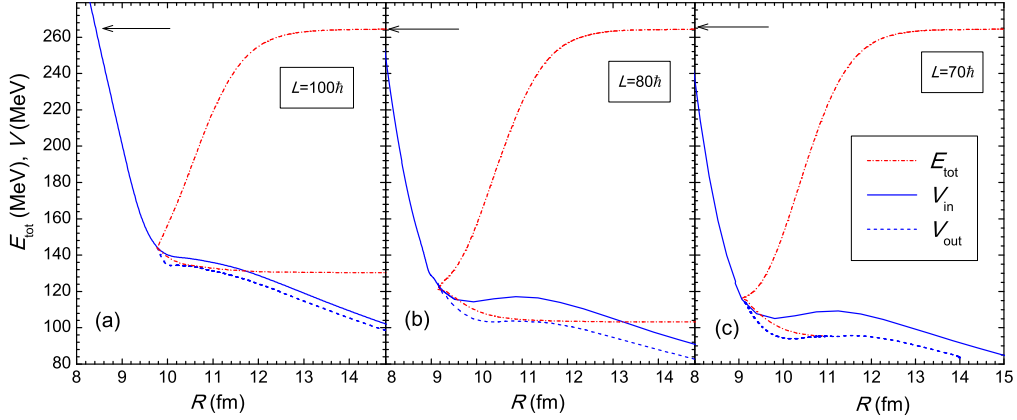}
\vspace*{-6.3 cm}
\caption{(Color online) 
Results of the dynamical calculations of the total energy $E_{\rm tot}$ (dot-dashed curves) and
the nucleus-nucleus interaction $V(R)$ (solid curves for the incoming path and dashed curves for the
outgoing path) as functions of the relative distance $R$ between the centres-of-mass of colliding
nuclei in the reaction of $\nuclide[78]{Kr} + \nuclide[40]{Ca}$.
The graphs (a) and (b) are examples of deep inelastic collisions with the dissipation of the kinetic
energy of the relative motion, while the graph (c) is one of the capture events when the system is trapped
into the potential well.
The nucleus-nucleus interaction in (b) contains the potential well while the one in (a) does not.
The presented results are obtained by the use of values
$\alpha_1^{} = 45^\circ$ and $\alpha_2^{} = 15^\circ$ of the orientation angles of the axial
symmetry of the nuclei relative to the beam direction.
The arrows show the points corresponding to the collision energy $E_{\rm c.m.} = 264~\text{MeV}$.}
\label{DicCap}
\end{figure}

The capture probability calculated through Eqs.~(\ref{maineq1})--(\ref{maineq5}) allows us to
determine the dissipation of the relative kinetic energy and angular momentum of the system.
The initial collision energy $E_{\rm c.m.}$ in the centre-of-mass system is shared by the kinetic energy
$E_{\rm kin}$ of the relative motion, nucleus-nucleus interaction $V(Z,\ell,R)$ and the dissipated energy
$E_{\rm diss}$ due to the radial and tangential friction forces, which leads to
$E_{\rm c.m.} = E_{\rm kin} + V(Z,\ell,R) + E_{\rm diss}$.
In Fig.~\ref{DicCap} we present the results of the dynamical calculations of the nucleus-nucleus
interaction $V(Z,\ell,R)$ and the total energy $E_{\rm tot} = E_{\rm kin} + V(Z,\ell,R)$ of the relative
motion which decreases due to dissipation, which show the difference between deep-inelastic collisions
[Fig.~\ref{DicCap}(a) and Fig.~\ref{DicCap}(b)] and the capture process with a full momentum transfer
[Fig.~\ref{DicCap}(c)] in the $\nuclide[78]{Kr} + \nuclide[40]{Ca}$ reaction.
The solid curves in Fig.~\ref{DicCap} show the values of $V(R)$ for the incoming path of collisions
and the dashed curves are obtained for the outgoing path as functions of the relative distance $R$
between the centres-of-mass of the colliding nuclei.
The graphs in Figs.~\ref{DicCap}(a) and \ref{DicCap}(b) are examples of deep inelastic collisions
with the dissipation of the kinetic energy of relative motion, while the graph in Fig.~\ref{DicCap}(c) is
one of the capture events when the system is trapped into the potential well.
These results show that the capture process does not take place in collisions with a large value of the
relative angular momentum, for example, at $L = 100 \hbar$, when there is no potential well as
illustrated in Fig.~\ref{DicCap}(a).
But the collision can be referred to as a deep-inelastic collision in the case of the presence of the potential
well if the dissipation of the relative kinetic energy cannot trap the system into the well as shown in
Fig.~\ref{DicCap}(b).
The collisions with $L \le 70 \hbar$ lead to capture processes as the total energy of DNS is trapped
into potential well  as in Fig.~\ref{DicCap}(c).

\begin{figure}[t]
\centering
\includegraphics[width=1.2\columnwidth]{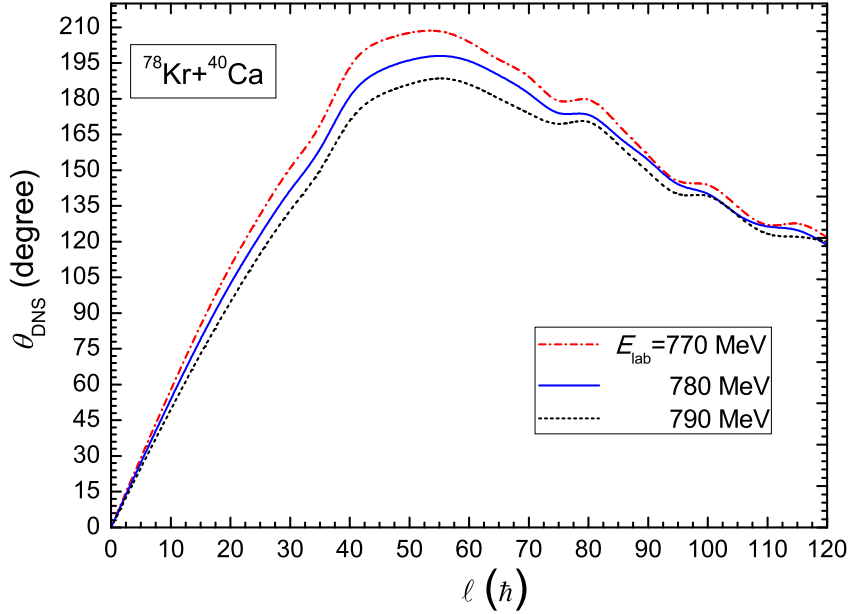}
\vspace*{-3. cm}
\caption{(Color online) 
The rotational angle of the dinuclear system formed in the $\nuclide[78]{Kr} + \nuclide[40]{Ca}$
reaction as a function of orbital angular momentum $\ell$ for a given $E_{\rm lab}$.
}
\label{KrCaRot}
\end{figure}

The angular distribution of the reaction products is obtained by calculating the rotational angle with the
lifetime and angular velocity of the DNS determined by Eq.~(\ref{tetaDNS}).
In Fig.~\ref{KrCaRot} we present the results for the rotational angle of the DNS formed in
the reaction of $\nuclide[78]{Kr \mbox{ ($10 A$ MeV})} + \nuclide[40]{Ca}$ as a function of
orbital angular momentum for several values of the initial energy.
It can be seen that, in the middle values ($40\hbar \mbox{ -- } 70\hbar$) of orbital angular momentum,
the rotational angle of the DNS is larger and the maximum value of the rotational angle is close to
$180^\circ$.
This means that the lifetime and rotational velocity of the DNS allow the projectile-like fragment to go
beyond the target fragment.
The smallness of the probability of the DNS decay in the perpendicular direction with
respect to the beam direction with $L = 40\hbar \mbox{ -- } 70\hbar$ may be understood from
Fig.~\ref{KrCaRot}.
Certainly the energy accumulated in the rotation of the DNS will increase the relative velocity of the decay
products in the forward and backward directions to the beam.
But the rotational energy contributing to the increase of the relative velocity of the decay products in the
perpendicular direction would be small since the corresponding values of the angular momentum are small
such as $L=10$---30$\hbar$ (see Fig. \ref{KrCaRot}).

This phenomenon is observed in heavy ion collisions even with massive nuclei.
For example, in Ref.~\cite{HCHH08}, the authors discussed the emission of the target-like
nucleus in the beam direction of the laboratory system but with a velocity smaller than that
of the compound nucleus.
The intensity of the low-velocity peak was found to be much lower than that of the high-velocity peak.
The two-peak structure was observed for all \nuclide{Rn}, \nuclide{Fr} and \nuclide{Ra} isotopes,
while it was found to fade for \nuclide{Po} and \nuclide{At}~\cite{HCHH08}.

Another example of the appearance of projectile-like products beyond target-like ones can be found
in the observation of the so-called ``slow'' evaporation residues in the reaction of
$\nuclide[20]{Ne} + \nuclide[208]{Pb}$ at projectile energies $E/A=8.6$ and $11.4$~MeV/u
reported in Ref.~\cite{HNHASV94},%
\footnote{We are grateful to the referee for informing us this work.}
where the velocity decrease was observed for the massive target-like products which are considered as the
evaporation residues being registered by the velocity filter SHIP at GSI.
At some values of orbital angular momentum $L$, the DNS formed after the capture of the
\nuclide[20]{Ne} nucleus by the \nuclide[208]{Pb} target nucleus can rotate around the axis going
through its centre-of-mass and breaks down into two fragments.
For the direction of the heavy (target-like) product velocity being parallel to the beam line, the rotational
angle of the DNS should be around $180^{\circ}$.
A simple calculation shows that, for example, the velocity of \nuclide[213]{Fr} is
${\rm v}_{\rm Fr}^{} = 0.18$~cm/ns after decay of the DNS which is formed at projectile energies
$E/A=8.6$~MeV/u.
At this initial energy the velocity of the centre-of-mass system is about ${\rm v}_{\rm CN}^{} =0.36$~cm/ns.
In the experiment discussed in Ref.~\cite{HNHASV94}, the maximum value of the velocity distribution of
the ``slow'' \nuclide[213]{Fr} isotope was found to be 
${\rm v}_{\rm Fr}^{}/{\rm v}_{\rm CN}^{} \simeq 0.5$.
This leads to the conclusion that the yield of ``slow'' evaporation residues observed in the
$\nuclide[20]{Ne} + \nuclide[208]{Pb}$ reaction at the projectile energy $E/A=8.6$~MeV/u
comes from quasifission process.
In the case of incomplete fusion, the velocity of the target-like reaction products is close to that of the
compound nucleus (${\rm v}_{\rm Fr}^{}/{\rm v}_{\rm CN}^{} \simeq 0.9$) whereas the differences between the velocities of quasifission products and
the compound nucleus can be noticeable as a function of the mass ratio of the binary fission-like products.

\begin{figure}[t]
\centering
\includegraphics[width=1.1\columnwidth]{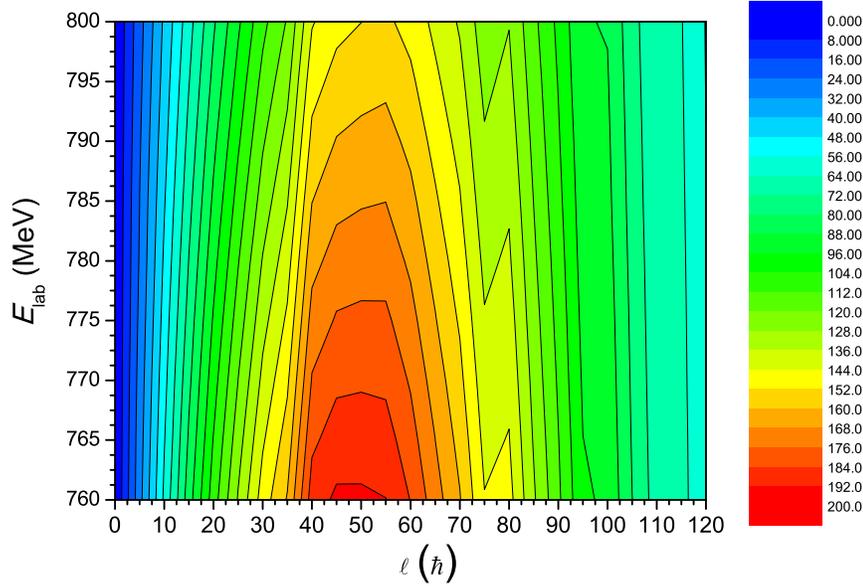}
\vspace*{-2.0 cm}
\caption{(Color online) 
Contour diagram for the rotational angle of the dinuclear system formed in the
$\nuclide[78]{Kr} + \nuclide[40]{Ca}$ reaction as a function of the angular momentum
$\ell$ and energy $E_{\rm lab}$.  }
\label{AnglQfis}
\end{figure}

Combined with Fig.~\ref{KrCaRot}, Fig.~\ref{AnglQfis} illustrates the dependence of the rotational angle
of the DNS formed in the $\nuclide[78]{Kr} + \nuclide[40]{Ca}$ reaction with angular momentum
$\ell$ for the initial energy from 760~MeV to 800~MeV.
It is found that, in the collision with the initial values of $L=50 \hbar$ and $E_{\rm lab}=770$~MeV,
the rotational angle of the DNS has the maximum value that corresponds to the situation when the
projectile and target nuclei exchange their positions relative to the beam direction.
Then, after the decay of the DNS, the projectile-like product can be observed in the forward hemisphere
with a speed larger than that of the compound nucleus due to the repulsion by the Coulomb
force of the target-like products.
This phenomenon is consistent with the observation discussed in Ref.~\cite{ISODEC-14}.
The relative velocity of these fragments is in the range of 2.4 -- 2.7~cm/ns, which overlaps with the
experimental data presented in Fig.~2 of Ref.~\cite{ISODEC-14}, where the yield of binary fragments
flying in the opposite direction, i.e., $-1.0 < \cos(\alpha) < -0.7$ with $\alpha$ being the folding angle
between the centre-of-mass velocities of the two fragments was discussed.
This observation was interpreted in Ref.~\cite{ISODEC-14} as a new reaction mechanism of a prompt
shock-induced fission following the fusion of \nuclide[78]{Kr} and \nuclide[40]{Ca} nuclei.

Therefore, through the present work, we suggest another mechanism of quasifission producing
massive products in the forward hemisphere in capture reactions.
The products formed through this mechanism can contribute to the yield of the fragments observed in
Ref.~\cite{ISODEC-14}.
We also find that the rotational velocities of the reaction products around their own axes are very small,
and certainly the alpha particles emitted from these products after quasifission are
expected to be distributed isotropically if the intrinsic spin of the product which emits
$\alpha$ particles is small.

\begin{figure}[t]
\centering
\includegraphics[width=1.0\columnwidth]{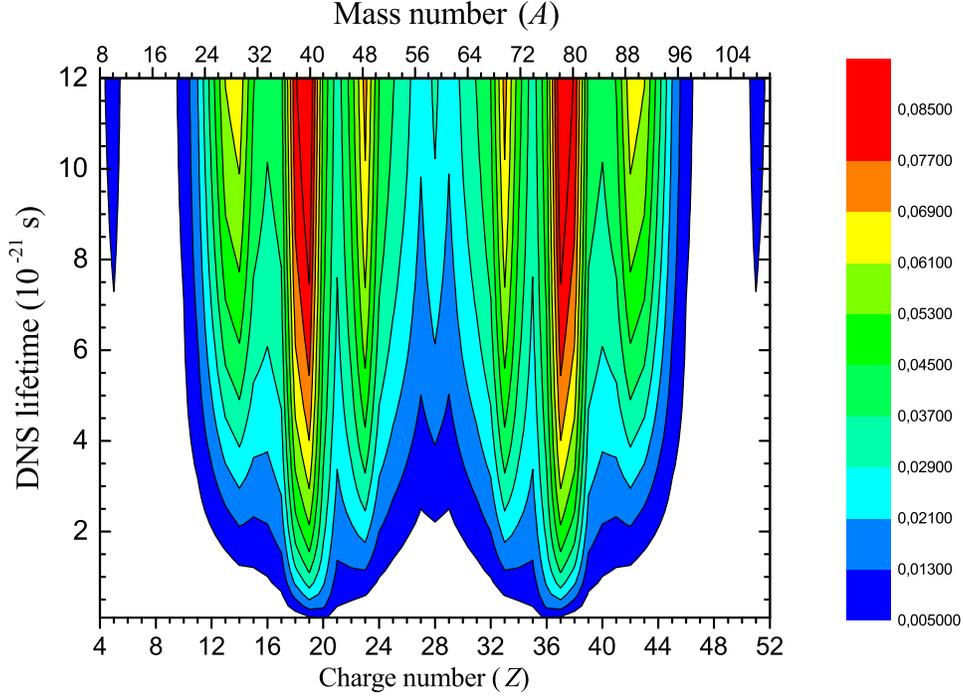}
\vspace*{-0.5 cm}
\caption{(Color online) 
Evolution of the charge distribution of the quasifission products as a function of the lifetime of
the DNS formed in the $\nuclide[78]{Kr} + \nuclide[40]{Ca}$ reaction at the beam energy of
$E_{\rm lab} = 10~\mbox{MeV}/A$.
The mass numbers shown on the top axis of the figure correspond to the primary products of the
reaction.}
\label{KrCaYield}
\end{figure}

\begin{figure}[t]
\centering
\includegraphics[width=1.\columnwidth]{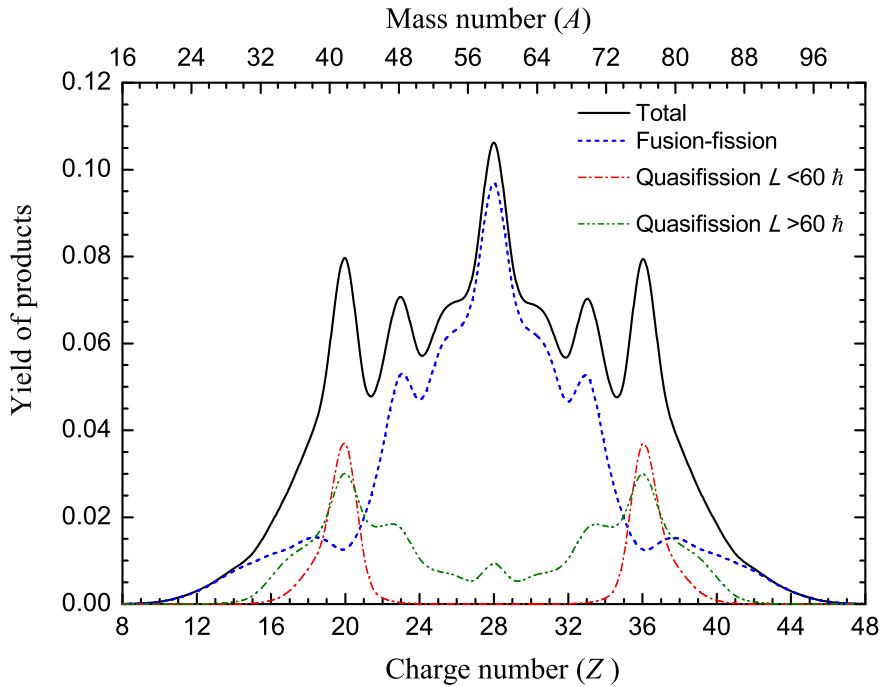}
\vspace*{-0.75 cm}
\caption{(Color online) 
The charge (mass) distribution of the quasifission (dot-dashed and dot-dot-dashed curves)
and fusion-fission (dashed curve) products calculated for the $\nuclide[78]{Kr} + \nuclide[40]{Ca}$
reaction at the beam energy of $E_{\rm lab} = 10~\mbox{MeV}/A$.
The total yield is shown by the solid line.
The mass numbers shown on the top axis correspond to that of the primary products of the reaction.}
\label{MixedYield}
\end{figure}

It is clearly seen in Figs.~\ref{KrCaYield}  and \ref{MixedYield} that in collisions with $L< 60 \hbar$
the centroids of the charge and mass distributions of the quasifission products concentrate at around
$Z_L = 18$ and $A_L = 38$ for the light product and around $Z_H = 38$
and $A_L = 78$ for the heavy product.
The mass numbers shown on the top axis of Figs.~\ref{KrCaYield} and \ref{MixedYield} correspond
to those of the primary products of the reaction.
The shape of the charge distribution is the manifestation of nuclear shell effects related with the
closed shells with the neutron numbers $N = 20$ and $40$.
The shell effects in the theoretical curves of the the charge distribution of primary products survive
due to accumulation of the part of the collision energy in the rotational degrees of freedom
(about 40~MeV) and direct dependence of the transition coefficients on the single-particle energies
of nucleons in the DNS nuclei.
The gaps between energy levels in light nuclei are larger than those in massive nuclei and this promotes
the appearance of the shell effects.
The shape of the charge and mass distributions of the quasifission process depends on the orbital
angular momentum.
In collisions with $60 \hbar < L < 70\hbar$ the charge and mass distributions
extend up to the mass symmetric region by overlapping with those of the fusion-fission products.

The results presented in Figs.~\ref{AnglQfis} and \ref{KrCaYield} for the angular and mass
distributions, respectively, of the quasifission products show their contribution to anisotropy of the velocity
distribution of the fission-like products observed in the experiment of Refs.~\cite{ISODEC-14,Schroeder15}.
This shows that the model considered in the present work can be considered as an alternative
interpretation of the new mode of the prompt fission of the composite nucleus formed in the reaction of
$\nuclide[78]{Kr \mbox{ (10 $A$ MeV)}} + \nuclide[40]{Ca}$.
This means that the small elongation of the velocity distribution of the fission-like products reported
in Refs.~\cite{ISODEC-14,Schroeder15} might be related with the contribution of the
quasifission products.


\section{Summary and Conclusion}

In summary, we performed a theoretical study on the angular and mass distributions of
quasifission fragments in the reaction of $\nuclide[78]{Kr \mbox{ (10 $A$ MeV)}} + \nuclide[40]{Ca}$,
which indicates that they are wide and the rotational angle of the dinuclear system can reach $180^\circ$
at collisions with relative angular momentum of $L = \mbox{(40--60)}\hbar$.
As a result, the projectile-like products can be observed in the forward hemisphere with a velocity
in the range of $2.4$--$2.7$~cm/ns, which is consistent with the experimental observations
reported in Refs.~\cite{ISODEC-14,Schroeder15}.

The shape of the charge and mass distributions of the quasifission process depends on the orbital
angular momentum.
In collisions with $L< 60 \hbar$ the average values of the charge and mass distributions are
rather concentrated near the projectile/target masses and charges at around
($Z_L = 18$, $A_L = 38$) for the lighter product and at around ($Z_H = 38$, $A_H = 78$)
for the heavier product.
In collisions with $60 \hbar < L < 80\hbar$ the charge and mass distributions extend up to the
mass symmetric region overlapping with those of the fusion-fission products.

In the experiment of Ref.~\cite{ISODEC-14}, the emission of alpha-particles was also found to be nearly
isotropic being emitted from the projectile-like products in the forward hemisphere.
In the present work, we found that the quasifission mechanism can reproduce the observed
angular and mass distributions of these projectile-like products.
The energy accumulated due to the rotation of the DNS increases the relative velocity of the decay
products in the forward and backward directions since a relatively large value of angular momentum,
namely, $L = (40 \mbox{ -- } 70)\hbar$, allows DNS decays in these directions.
The rotational energy contributing to the increase of the relative velocity of the decay products
in the perpendicular direction is, however, small due to the small value of the corresponding
angular momentum, $L = (10 \mbox{ -- } 30) \hbar$.
As a result, the velocity distribution of the fission-like products observed in the experiment of
Refs.~\cite{ISODEC-14,Schroeder15} can have a slightly elongated shape along the beam direction.

\section*{Acknowledgements}

A.K.N. was supported in part by the Russian Foundation for Basic Research
under Grant No. 10-02-0030 and the Academy of Science of Uzbekistan under 
Grant No.\ F2-FA-F115.
He also acknowledges the support from the MSIP of the Korean Government under the Brain Pool
Program No.\ 142S-1-3-1034.
The work of Y.O.  was supported by the Basic Science Research Program through the National
Research Foundation of Korea (NRF) under Grant No. \\ NRF-2013R1A1A2A10007294.

\end{document}